# Revisiting the buckling metrology method to determine the Young's modulus of 2D materials

*Nestor Iguiñiz[1], Riccardo Frisenda[1], Rudolf Bratschitsch[2], Andres Castellanos-Gomez[1],\**

[1] *Materials Science Factory, Instituto de Ciencia de Materiales de Madrid (ICMM), Consejo Superior de Investigaciones Científicas (CSIC), Sor Juana Inés de la Cruz 3, 28049 Madrid, Spain.*
[2] *Institute of Physics and Center for Nanotechnology, University of Münster, 48149 Münster, Germany.*
E-mail: andres.castellanos@csic.es

Measuring the mechanical properties of two-dimensional materials is a formidable task. While regular electrical and optical probing techniques are suitable even for atomically thin materials, conventional mechanical tests cannot be directly applied. Therefore, new mechanical testing techniques need to be developed. Up to now, the most widespread approaches require micro-fabrication to create freely suspended membranes, rendering their implementation complex and costly. Here, we revisit a simple yet powerful technique to measure the mechanical properties of thin films. The buckling metrology method, that does not require the fabrication of freely suspended structures, is used to determine the Young's modulus of several transition metal dichalcogenides ($MoS_2$, $MoSe_2$, $WS_2$ and $WSe_2$) with thicknesses ranging from 3 to 10 layers. We critically compare the obtained values for the Young's modulus and their uncertainty, finding that this simple technique provides results, which are in good agreement with those reported using other highly sophisticated testing methods. By comparing the cost, complexity and time required for the different methods reported in the literature, the buckling metrology method presents certain advantages that makes it an interesting mechanical test tool for 2D materials.

Two-dimensional (2D) materials are promising candidates for future flexible electronics applications due to their combination of remarkable mechanical and electrical properties.[1] In fact, from the mechanics point of view, 2D materials are similar to polymers, as they are very elastic and resilient to large deformations [2,3], while keeping electronic performance comparable to that of crystalline 3D materials.[4,5]

While the electrical and optical properties of 2D materials can be explored with conventional experimental tools developed to test 3D materials and thin film devices, probing the mechanical properties of 2D materials is more challenging, as neither bending nor tensile test macroscopic setups can be employed. Nanoindentation,[2,6–8] the analysis of the dynamics of nanomechanical resonators,[9] and the microscopic adaptation of tensile tests setups[10] or Brillouin scattering[11] have been developed to characterize the fundamental mechanical properties of 2D materials such as their Young's modulus.[12,13] Although powerful, these techniques require dedicated setups and rather complex data acquisition and/or analysis. Alternative to these methods, Stafford *et al.*[14] introduced the buckling metrology method, a simple and elegant way to measure the Young's modulus of thin polymeric films by studying the buckling instability, which arises when the film is deposited onto a compliant substrate, and it is subjected to uniaxial compression.[15] Under these conditions, the trade-off between the adhesion forces between film and substrate and the bending rigidity of the film leads to a rippling of the thin film with a characteristic wavelength that only depends on the elastic properties of the film and the substrate. This elegant method to characterize the mechanical properties of thin films has been extensively used to study coatings [14] and organic semiconducting materials.[16] However, it has been scarcely employed to study 2D materials[17–20], and it seems that is has been mostly overlooked by the 2D materials community.

Here, we apply the buckling-based metrology method to determine the Young's modulus of transition metal dichalcogenide flakes with thickness ranging from 3 layers up to 10 layers. We use optical microscopy to determine both the number of layers and the rippling wavelength, which is therefore very fast and simple to implement. We critically compare the results obtained with this method and demonstrate that despite its simplicity it provides results in good agreement with other techniques to study the mechanical properties of 2D materials. We believe that the buckling-based metrology method provides a fast route to determine the Young's modulus of 2D materials, being an excellent alternative to other existing nanomechanical test methods that are more technically demanding.

The samples are fabricated by mechanical exfoliation of bulk layered TMDC crystals with adhesive tape (see Experimental section for details). The exfoliated material is then transferred onto a compliant elastomeric substrate (Gelfilm®, a commercially available polydimethylsiloxane, PDMS, film manufactured by Gelpac®) which is subjected to an uniaxial stress of ~20%. Right after the transfer, the stress on the elastomeric substrate is released, yielding to compressive uniaxial strain of the transferred flakes (**Figure 1**a shows a schematic diagram of the sample fabrication process). Note that due to the large Young's modulus mismatch between the elastomeric substrate and the 2D materials only a small fraction of the substrate pre-stress will be transferred to the flakes after releasing the stress. Figure 1b displays optical microscopy images of a multilayered $MoS_2$ flake after being transferred onto the pre-stressed elastomeric substrate and right after releasing the stress on the substrate.

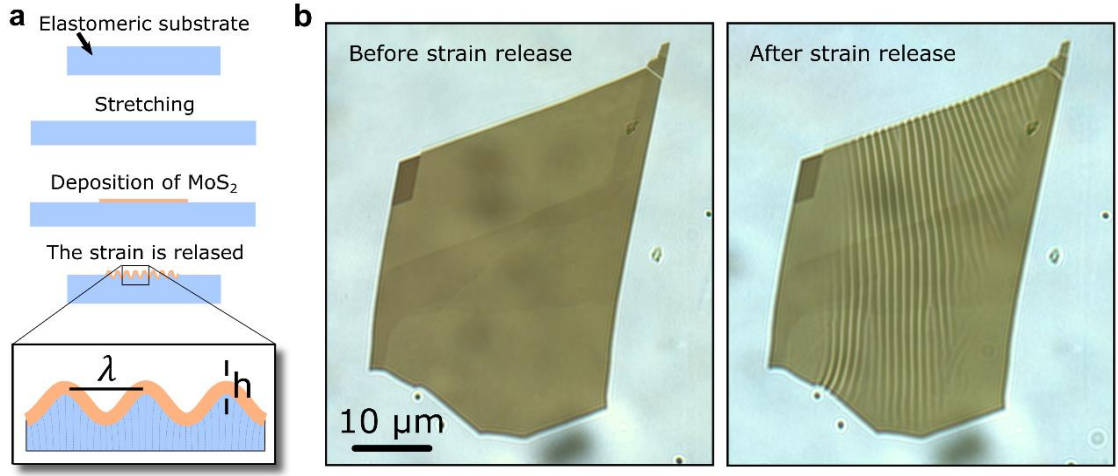

**Figure 1.** (a) Sketch of the process employed to fabricate the samples used in the buckling metrology method. The flakes are transferred onto a stressed elastomeric substrate, when the stress on the substrate is released the flakes are subjected to uniaxial compressive strain that produces the buckling of the flakes. (b) Transmission mode optical microscopy images of a MoS$_2$ multilayer flake (the thinner region is 7L thick) before and after releasing the stress on the elastomeric substrate.

The rippled pattern (that can be seen more easily in the transmission mode optical image) arises from the buckling instability resulting from the balance between the energy required to bend the stiff 2D material, the energy to elastically deform the soft underlying Gelfilm and the adhesion energy between them. Interestingly, the wavelength of these ripples is independent on the initial pre-stress of the elastomeric substrate and it only depends on the materials properties of both flake and substrate:[14,21,22]

$$\lambda = 2\pi h \left[\frac{(1-v_s^2)E_f}{3(1-v_f^2)E_s}\right]^{1/3} \quad (1)$$

where $h$ is the flake thickness, $v_s$ and $v_f$ are the Poisson's ratio of substrate and flake and $E_s$ and $E_f$ are the Young's modulus of the substrate and flake respectively. This equation is valid under certain assumptions: (a) the flake should follow a sinusoidal rippling, (b) $E_f/E_s \gg 1$, (c) the substrate should be much thicker than the flake, (d) the amplitude of the ripples should be much smaller than their wavelength (thus the shear forces are neglected), (e) the adhesion between the flake and substrate is strong enough to prevent slippage and (f) all the deformations are assumed to be elastic.

According to Equation [1] spatial wavelength of the ripples monotonically depends on the thickness of the flakes, as the other parameters are fixed and they only depend on the intrinsic mechanical properties of the substrate and flake materials. **Figure 2** shows a comparison between the rippled pattern observed in 3L, 7L and 10L $MoS_2$ flakes whose thickness is determined via quantitative analysis of their transmittance [23] (see the Supporting Information). The optical microscopy-based thickness determination method provides accurate thickness values for flakes in the 1 to 10 layers range. For thicker flakes the absorption starts to saturate and thus the thickness uncertainty rapidly increases.

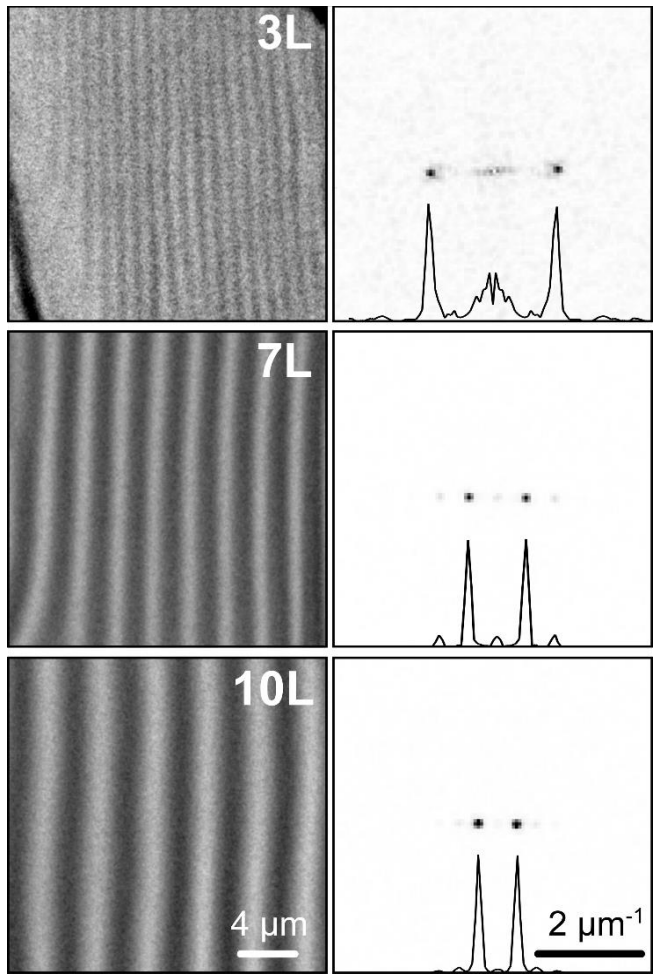

**Figure 2**. Grayscale white-light transmission mode optical microscopy images of the rippled pattern observed in 3L, 7L and 10L MoS$_2$ flakes and their corresponding FFTs. Line cuts along the FFT maxima are included in the FFT panels.

Using Equation [1], the Young's modulus of the deposited 2D materials can be determined by measuring the thickness-dependent ripple period provided that $\nu_s$, $\nu_f$ and $E_s$ are known values. The Poisson's ratio of the Gelfilm (PDMS) substrate $\nu_s = 0.5$ [24] and flake $\nu_f = 0.27$ [25,26] are found in the literature, with low spread in their values. Moreover, the Young's modulus is rather insensitive to small variations of the Poisson's ratio (see Supporting Information Figure S11). Because the PDMS Young's modulus values given in the literature show a large scattering (from 300 to 1000 kPa, strongly dependent on the curing process)[27,28], we have experimentally determined the Young's modulus of our Gelfilm substrate $E_s = 492 \pm 11$ kPa (see the Experimental section and the Supporting Information for more details about the Gelfilm Young's modulus determination). The Young's modulus of MoS$_2$ can be determined from the slope of the linear relationship between the ripple period ($\lambda$) and the flake thickness ($h$) as follows:

$$E_f = \frac{3(1-v_f^2)E_s}{8\pi^3(1-v_s^2)}\left[\frac{\lambda}{h}\right]^3 \quad (2)$$

**Figure 3**a displays the measured ripple period and the flake thickness, determined from the analysis of the transmission mode optical images, of flakes 3 to 10 layers thick. Note that flakes thinner than 3 layers present ripples with a period < 0.6 µm and a small amplitude (< 6 nm) that cannot be resolved with optical microscopy (see the Supp. Info.) and flakes thicker than ~10 layers typically yield very large flake-to-flake variation of their mechanical properties which has been attributed to the presence of stacking faults.[12,29,30] In order to resolve the ripples in flakes thinner than 3 layers one can alternatively use AFM (see some experimental datapoints acquired by measuring the ripple wavelength through AFM in Figure 3a and its associated discussion in the Supp. Info.). The experimental data follows a clear linear trend, in agreement with Equation [1] and from its slope we determine $E_f$ following Equation [2]: $E_f$ = 246 ± 35 GPa (see the Supporting Information for a discussion about the uncertainty determination). We observed a relatively large flake-to-flake variation of the ripple period (i.e. flakes with same thickness yield sizeably different ripple periods), which we attribute to the presence of small defects such as folds or wrinkles in some of the flakes. Therefore, measurements of several flakes with different thicknesses is needed to obtain a well-defined Young's modulus value with a low uncertainty. We found that measurements over at least 6-8 flakes (with ≥4 different thicknesses) are needed to determine the Young's modulus with an uncertainty comparable to that obtained with other mechanical testing techniques like nanoindentation (see the Supporting Information, Figure S9 and Table S3).

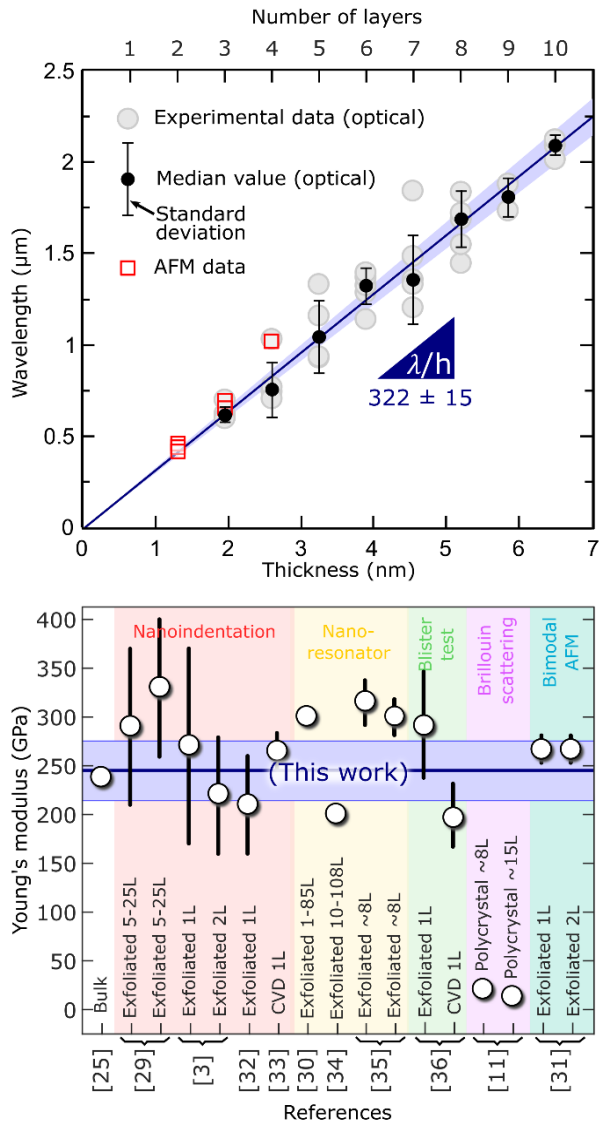

**Figure 3**. (a) Relationship between the wavelength of the ripples and the thickness of the MoS$_2$ flakes. The solid dark blue line represents a linear fit to the experimental data, the shaded light blue area around it indicates the uncertainty of the fit. The slope of the wavelength vs. thickness relationship, from which the Young's modulus of MoS$_2$ can be determined, is also included in the plot. (b) Summary of the values for the MoS$_2$ Young's modulus, reported in in the literature and their comparison with the value obtained in this work (solid dark blue line, the shaded light blue area indicates the uncertainty) E = 246 ± 35 GPa. Details about the sample fabrication and the measurement technique for the reference datapoints have been included in the plot.

This Young's modulus value is in good agreement with the values reported in the literature, determined through different experimental techniques. Figure 3b shows a comparison between the different values of the Young's modulus for few-layer $MoS_2$ available in the literature. In the Figure we specified the testing method and the sample fabrication method. The value determined with the buckling metrology method is in good agreement with the bulk value of $MoS_2$, measured by neutron dispersion and X-ray diffraction measurements, and it is compatible within experimental uncertainties with most of recent works that studied ultrathin flakes through nanoindentation, the analysis of the dynamics of nanomechanical resonators, the microscopic version of the blister test and bimodal atomic force microscopy.[3,11,36,25,29–35] The only noticeable disagreement is with the values reported in Ref. [[11]], measured with a microscopic version of Brillouin light scattering. In that reference, however, the authors measured a polycrystalline sample fabricated by direct sulfurization of a metallic film. All the information contained in Figure 3b is also summarized in Table 1 to facilitate a quantitative comparison between the different methods.

**Table 1** summarizes values reported in the literature for other TMDCs ($MoSe_2$, $WS_2$ and $WSe_2$)[32,37,38] and directly compares these literature values with the ones obtained by repeating the process described above for $MoS_2$ for the other members of the transition metal dichalcogenide family (see the Supporting Information). From the comparison between the S- and Se-based TMDCs there is a clear trend: the Young's modulus of the S-based TMDCs is larger than that of Se-based ones. This trend is in good agreement with ab initio calculations.[26,39]

**Table 1.** Summary of values for the Young's modulus (and their uncertainties) reported in the literature for $MoS_2$, $MoSe_2$, $WS_2$ and $WSe_2$. The testing method as well as some characteristics

of the studied samples (fabrication method and number of layers) are highlighted in the table to facilitate the comparison between the different reported values.

| Material | Testing method | Sample | | E (GPa) | Reference |
|---|---|---|---|---|---|
| | | Isolation method | # of layers | | |
| MoS$_2$ | Brillouin scattering | Bulk natural crystal | | 238 | [25] |
| | Nanoindentation | Exfoliation (natural) | 5 to 25 | 290 ± 80 | [29] |
| | | | 5 to 25 | 330 ± 70 | [29] |
| | | | 1 | 270 ± 100 | [3] |
| | | | 2 | 220 ± 60 | [3] |
| | | | 1 | 210 ± 50 | [31] |
| | | CVD | 1 to 2 | 264 ± 18 | [32] |
| | Dynamics of mechanical resonator | Exfoliation (natural) | 1 to 85 | 300 | [30] |
| | | | 10 to 108 | 200 | [33] |
| | | | ~8 | 315 ± 23 | [34] |
| | | | ~8 | 300 ± 18 | [34] |
| | Micro-Buggle test | Exfoliation (natural) | 1 | 292 ± 54 | [35] |
| | | CVD | 1 | 197 ± 31 | [35] |
| | micro-Brillouin light scattering | CVD-poly | ~8 | 20.1 ± 3.2 | [11] |
| | | | ~15 | 13.7 ± 2 | [11] |
| | Bimodal AFM | Exfoliation (natural) | 1 | 265 ± 13 | [36] |
| | | | 2 | 265 ± 13 | [36] |
| | **Buckling metrology method** | **Exfoliation (natural)** | **3 to 11** | **246 ± 35** | **This work** |
| MoSe$_2$ | Micro-tensile test | CVD | 1 to 2 | 177.2 ± 9.3 | [37] |
| | **Buckling metrology method** | **Exfoliation (synthetic)** | **5 to 10** | **224 ± 41** | **This work** |
| WS$_2$ | Nanoindentation | CVD | 1 | 272 ± 18 | [32] |
| | **Buckling metrology method** | **Exfoliation (synthetic)** | **3 to 8** | **236 ± 65** | **This work** |
| WSe$_2$ | Nanoindentation | Exfoliation (synthetic) | 5 to 12 | 167 ± 7 | [38] |
| | **Buckling metrology method** | **Exfoliation (synthetic)** | **4 to 9** | **163 ± 39** | **This work** |

It is important to make a critical comparison between the different experimental techniques employed to determine the Young's modulus values of 2D materials. In **Table 2** we present key information about the requirements needed by these different methods for the sample fabrication and the measurement process as well as a coarse estimation of the time needed for those processes. The first noticeable difference is that unlike the other methods, the buckling metrology method does not require any lithographic process. On the other hand, the nanoindentation, the nanomechanical resonator-based method, the micro-blister test, the microscopic version of the tensile test and the micro-Brillouin light scattering requires to fabricate freely suspended nanosheets, which significantly increases the complexity of the fabrication process and the time involved and it decreases the success yield. Moreover, we stress that the requirement of expensive lithography and etching setups necessary to fabricate the suspended sheets might be highly restrictive by preventing the implementation of these methods in many research groups without access to cleanroom facilities. The nanomechanical resonator-based method also requires the use of a specialized experimental setup for the measurement of the mechanical properties of the flakes, which again can be a handicap for its implementation in many research groups. The micro-Brillouin scattering method also needs a very specialized equipment.[11] The sample fabrication of the buckling metrology method, on the other hand, does not require any specialized technique as it simply relies on the exfoliation of the 2D materials on top of the stressed elastomer substrate and an optical microscope. Common to all the methods is the need to determine the number of layers, which can be done with atomic force microscopy (time needed for the measurement ~30-60 min) or optical microscopy (~1-10 min).

**Table 2.** Critical comparison between the different methods reported in the literature to test the mechanical properties of 2D materials. Key requirements for the sample fabrication and the measurement process are indicated as well as a qualitative estimation of the time needed in these processes and the complexity of the implementation (indicated with the symbol "●", the more symbols the more time/complexity is needed). *Multiple flakes are needed in the buckling metrology method to reduce the experimental uncertainty (see the main text).

| Technique | Sample fabrication | | Measurement | | Complexity | Cost | Ref. |
|---|---|---|---|---|---|---|---|
| | Requirements | Time | Requirements | Time | | | |
| Nanoindentation | Lithography + flake transfer | ●● | AFM<br>Air atmosphere | ●● | ●● | ●● | [2,3,29,31,32] |
| Nanomechanical resonator | Lithography + flake transfer | ●● | Interferometer (or electrical read-out)<br>High frequency electronics<br>High vacuum | ●● | ●●● | ●● | [30,33,34] |
| Micro-blister test | Lithography + flake transfer + high pressure chamber | ●●● | AFM<br>Air atmosphere | ●●● | ●●● | ●● | [35] |
| Micro-Brillouin scattering | Lithography + flake transfer | ●● | Modified micro-Raman spectrometer<br>Air atmosphere | ●● | ●● | ●● | [11] |
| Micro-tensile test | Lithography + flake transfer | ●●● | SEM<br>High vacuum | ●● | ●●● | ●●● | [37] |
| Bimodal AFM | Direct exfoliation on substrate | ● | AFM<br>Air atmosphere | ● | ●● | ●● | [36] |
| **Buckling metrology method** | **Direct exfoliation on substrate** | ● | **Optical microscope<br>Air atmosphere<br>Multiple flakes*** | ●●* | ● | ● | **This work** |

CONCLUSIONS

In summary, the buckling metrology method provides a fast and easy way of measuring the mechanical properties of 2D materials as compared with conventionally employed approaches (nanoindentation, nanomechanical resonators, blister test and micro-Brillouin light scattering). We demonstrate this method with $MoS_2$ and found that it provides Young's modulus values in good agreement with the literature values. Because of its simplicity, the fast measurement speed and straightforwardness of the data analysis we believe that this method can be a highly attractive way to study the mechanical properties of 2D materials.

**NOTE:** During the elaboration of this manuscript we became aware of a recent publication where another technique to measure the mechanical properties of supported (not freely suspended) 2D materials.[31] The mechanical properties of single- and bilayer $MoS_2$ supported onto a $SiO_2$/Si surface were measured with a bimodal AFM. The technique relies in the use of a multifrequency AFM in combination with finite elements analysis in order to extract Young's modulus.

**Experimental Section**

*Materials:*

$MoS_2$ samples were prepared out of a bulk natural molybdenite crystal (Moly Hill mine, Quebec, Canada). $MoSe_2$ and $WSe_2$ samples were prepared out of bulk synthetic crystals grown by physical vapour transport method (provided by Prof. Rudolf Bratschitsch). $WS_2$ samples were prepared out of a bulk synthetic crystal grown by physical vapour transport method at Tennessee Crystal Center. The elastomer substrate used in this work is a commercially available polydimethylsiloxane-based substrate manufactured by Gelpak® (both Gelfilm® WF X4 6.0 mil and Gelfilm® PF X4 6.5 mil were used with identical results).

*Determination of the Young's modulus of the Gelfilm® substrate:*

The Young's modulus of the elastomeric substrate has been determined by means of a force vs. elongation experiment where different forces are applied to a Gelfilm® strip (63.5 mm × 10 mm × 0.165 mm) by loading different test masses at one end of the strip and monitoring the relative changes in length upon loading though a Canon EOS 1200D camera equipped with a EF-S 18-55 DC III objective lens. See the Supporting Information for more details.

*Optical microscopy:*

Optical microscopy images have been acquired with an AM Scope BA MET310-T upright metallurgical microscope equipped with an AM Scope mu1803 camera with 18 megapixels. The calibration of the optical magnification system has been carried out by imaging standard samples: one CD, one DVD, one DVD-R, and two diffraction gratings with 300 lines/mm (Thorlabs GR13-0305) and 600 lines/mm (Thorlabs GR13-0605). See details about the calibration in the Supporting Information.

*Image analysis:*

The quantitative analysis of the transmittance of the flakes and the rippling wavelength has been carried out using Gwyddion® software.[39]

*Thickness determination:*

The thickness determination has been carried out by extracting the transmittance of the blue channel of the transmission mode optical microscopy images and comparing it with the results of a reference (not-buckled) sample. See the Supporting Information for more details.

*Atomic Force Microscopy (AFM):*

Atomic force microscopy measurements were carrier out with an ezAFM from NanoMagnetics Instruments operated in tapping mode with cantilevers of 40 N/m and a resonance frequency of 300 kHz.

**Acknowledgements**

This project has received funding from the European Research Council (ERC) under the European Union's Horizon 2020 research and innovation programme (grant agreement n° 755655, ERC-StG 2017 project 2D-TOPSENSE).

**References**

[1] D. Akinwande, N. Petrone, J. Hone, *Nat. Commun.* **2014**, *5*, 5678.

[2] C. Lee, X. Wei, J. W. Kysar, J. Hone, *Science* **2008**, *321*, 385.

[3] S. Bertolazzi, J. Brivio, A. Kis, *ACS Nano* **2011**, *5*, 9703.

[4] B. Radisavljevic, A. Radenovic, J. Brivio, V. Giacometti, A. Kis, *Nat. Nanotechnol.* **2011**, *6*, 147.

[5] Q. H. Wang, K. Kalantar-Zadeh, A. Kis, J. N. Coleman, M. S. Strano, *Nat. Nanotechnol.* **2012**, *7*, 699.

[6] M. Poot, H. S. J. van der Zant, *Appl. Phys. Lett.* **2008**, *92*, 63111.

[7] C. Gómez-Navarro, M. Burghard, K. Kern, *Nano Lett.* **2008**, *8*, 2045.

[8] I. W. Frank, D. M. Tanenbaum, A. M. van der Zande, P. L. McEuen, *J. Vac. Sci. Technol. B Microelectron. Nanom. Struct. Process. Meas. Phenom.* **2007**, *25*, 2558.

[9] J. S. Bunch, A. M. van der Zande, S. S. Verbridge, I. W. Frank, D. M. Tanenbaum, J. M. Parpia, H. G. Craighead, P. L. McEuen, *Science* **2007**, *315*, 490.

[10] Y. Yang, X. Li, M. Wen, E. Hacopian, W. Chen, Y. Gong, J. Zhang, B. Li, W. Zhou, P. M. Ajayan, Q. Chen, T. Zhu, J. Lou, *Adv. Mater.* **2017**, *29*, 1604201.

[11] B. Graczykowski, M. Sledzinska, M. Placidi, D. Saleta Reig, M. Kasprzak, F. Alzina, C. M. Sotomayor Torres, *Nano Lett.* **2017**, *17*, 7647.

[12] A. Castellanos-Gomez, V. Singh, H. S. J. van der Zant, G. A. Steele, *Ann. Phys.* **2015**, *527*, 27.

[13] X. Li, M. Sun, C. Shan, Q. Chen, X. Wei, *Adv. Mater. Interfaces* **2018**, *5*, 1701246.

[14] C. M. Stafford, C. Harrison, K. L. Beers, A. Karim, E. J. Amis, M. R. VanLandingham, H.-C. Kim, W. Volksen, R. D. Miller, E. E. Simonyi, *Nat. Mater.* **2004**, *3*, 545.

[15] N. Bowden, S. Brittain, A. G. Evans, J. W. Hutchinson, G. M. Whitesides, *Nature*

**1998**, *393*, 146.

[16] M. A. Reyes-Martinez, A. Ramasubramaniam, A. L. Briseno, A. J. Crosby, *Adv. Mater.* **2012**, *24*, 5548.

[17] P. Feicht, R. Siegel, H. Thurn, J. W. Neubauer, M. Seuss, T. Szabó, A. V Talyzin, C. E. Halbig, S. Eigler, D. A. Kunz, *Carbon N. Y.* **2017**, *114*, 700.

[18] D. A. Kunz, J. Erath, D. Kluge, H. Thurn, B. Putz, A. Fery, J. Breu, *ACS Appl. Mater. Interfaces* **2013**, *5*, 5851.

[19] D. A. Kunz, P. Feicht, S. Gödrich, H. Thurn, G. Papastavrou, A. Fery, J. Breu, *Adv. Mater.* **2013**, *25*, 1337.

[20] C. J. Brennan, J. Nguyen, E. T. Yu, N. Lu, *Adv. Mater. Interfaces* **2015**, *2*, 1500176.

[21] A. L. Volynskii, S. Bazhenov, O. V Lebedeva, N. F. Bakeev, *J. Mater. Sci.* **2000**, *35*, 547.

[22] D. Khang, J. A. Rogers, H. H. Lee, *Adv. Funct. Mater.* **2009**, *19*, 1526.

[23] Y. Niu, S. Gonzalez-Abad, R. Frisenda, P. Marauhn, M. Drüppel, P. Gant, R. Schmidt, N. Taghavi, D. Barcons, A. Molina-Mendoza, S. de Vasconcellos, R. Bratschitsch, D. Perez De Lara, M. Rohlfing, A. Castellanos-Gomez, *Nanomaterials* **2018**, *8*, 725.

[24] R. H. Pritchard, P. Lava, D. Debruyne, E. M. Terentjev, *Soft Matter* **2013**, *9*, 6037.

[25] J. L. Feldman, *J. Phys. Chem. Solids* **1976**, *37*, 1141.

[26] D. Çakır, F. M. Peeters, C. Sevik, *Appl. Phys. Lett.* **2014**, *104*, 203110.

[27] I. D. Johnston, D. K. McCluskey, C. K. L. Tan, M. C. Tracey, *J. Micromechanics Microengineering* **2014**, *24*, 35017.

[28] K. Khanafer, A. Duprey, M. Schlicht, R. Berguer, *Biomed. Microdevices* **2009**, *11*, 503.

[29] A. Castellanos-Gomez, M. Poot, G. A. Steele, H. S. J. van der Zant, N. Agraït, G.

Rubio-Bollinger, *Adv. Mater.* **2012**, *24*, 772.

[30]  A. Castellanos-Gomez, R. van Leeuwen, M. Buscema, H. S. J. van der Zant, G. A. Steele, W. J. Venstra, *Adv. Mater.* **2013**, *25*, 6719.

[31]  R. C. Cooper, C. Lee, C. A. Marianetti, X. Wei, J. Hone, J. W. Kysar, *Phys. Rev. B* **2013**, *87*, 035423.

[32]  K. Liu, Q. Yan, M. Chen, W. Fan, Y. Sun, J. Suh, D. Fu, S. Lee, J. Zhou, S. Tongay, *Nano Lett.* **2014**, *14*, 5097.

[33]  J. Lee, Z. Wang, K. He, J. Shan, P. X.-L. Feng, *ACS Nano* **2013**, *7*, 6086.

[34]  D. Davidovikj, F. Alijani, S. J. Cartamil-Bueno, H. S. J. Zant, M. Amabili, P. G. Steeneken, *Nat. Commun.* **2017**, *8*, 1253.

[35]  D. Lloyd, X. Liu, N. Boddeti, L. Cantley, R. Long, M. L. Dunn, J. S. Bunch, *Nano Lett.* **2017**, *17*, 5329.

[36]  Y. Li, C. Yu, Y. Gan, P. Jiang, J. Yu, Y. Ou, D.-F. Zou, C. Huang, J. Wang, T. Jia, *npj Comput. Mater.* **2018**, *4*, 49.

[37]  Y. Yang, X. Li, M. Wen, E. Hacopian, W. Chen, Y. Gong, J. Zhang, B. Li, W. Zhou, P. M. Ajayan, *Adv. Mater.* **2017**, *29*, 1604201.

[38]  R. Zhang, V. Koutsos, R. Cheung, *Appl. Phys. Lett.* **2016**, *108*, 42104.

[39]  Z. Fan, Z. Wei-Bing, T. Bi-Yu, *Chinese Phys. B* **2015**, *24*, 97103.

[40]  C. Lee, H. Yan, L. E. Brus, T. F. Heinz, Ḱ. J. Hone, S. Ryu, *ACS Nano* **2010**, *4*, 2695.

Supporting Information

# Revisiting the buckling metrology method to determine the Young's modulus of 2D materials

*Nestor Iguiñiz, Riccardo Frisenda, Rudolf Bratschitsch, Andres Castellanos-Gomez\**

**Determination of the Young's modulus of the Gelfilm substrate**

**Calibration of the optical microscope system**

**Thickness determination**

**Young's modulus determination for MoSe$_2$, WS$_2$ and WSe$_2$**

**Example of ripples below the microscope resolution**

**Sensitivity of the determination of the Young's modulus on the choice of Poisson's ratio values**

**Determination of the Young's modulus of the Gelfilm substrate**

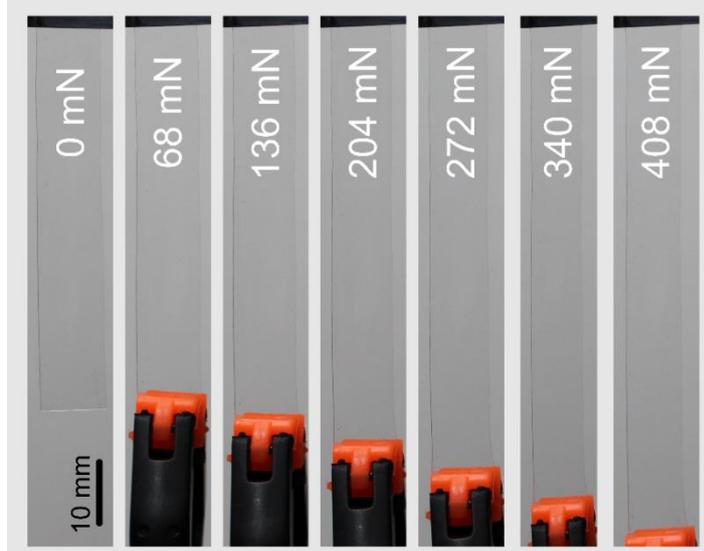

**Figure S1**. Gelfilm strip (10 mm wide, 165.1 µm thick) subjected to a tensile test. The load force is applied by using test masses. The sequence of optical images shows the change in length upon force load which can be used to quantitatively extract the Young's modulus of the Gelfilm.

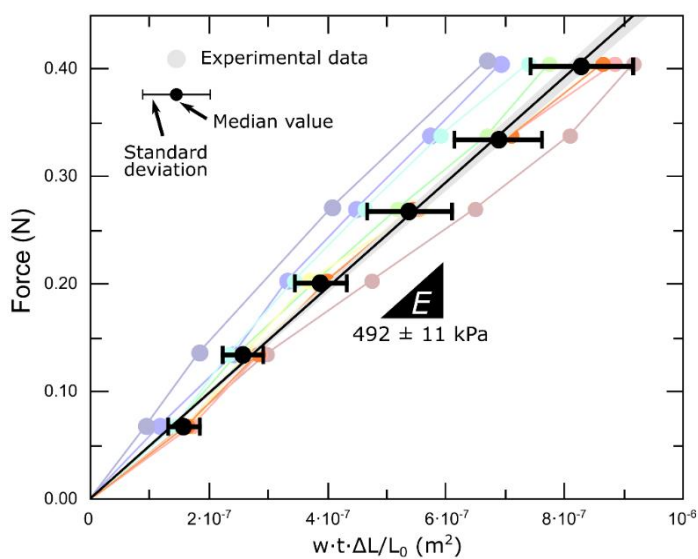

**Figure S2**. Force *vs*. elongation (ΔL) of the Gelfilm strips (9 different strips with different geometries, see Table S1, have been measured). The elongation has been normalized to the initial length and the width and thickness to compare datasets of strips with different geometries. This normalization also allows one to directly extract the Young's modulus from the slope of a linear fit to the experimental data. Note that for each force value the median normalized elongation value and the standard deviation have been calculated. The black solid line is the linear fit to the median value of the experimental data. The shaded light grey area indicates the uncertainty of the fit.

**Table S1**. Summary of the geometry of the different Gelfilm stripes studied in the force *vs*. elongation experiments employed to determine the Young's modulus of the Gelfilm.

| Sample | Type of gelfilm | Length (mm) | Width (mm) | Thickness (µm) |
|---|---|---|---|---|
| 1 | PF X4 6.5mil | 63.5 | 14.67 | 165.1 |
| 2 | PF X4 6.5mil | 63.5 | 10.25 | 165.1 |
| 3 | PF X4 6.5mil | 63.5 | 10.0 | 165.1 |
| 4 | PF X4 6.5mil | 63.5 | 9.86 | 165.1 |
| 5 | PF X4 6.5mil | 63.5 | 10.86 | 165.1 |
| 6 | PF X4 6.5mil | 63.5 | 10.0 | 165.1 |
| 7 | PF X4 6.5mil | 63.5 | 10.0 | 165.1 |
| 8 | PF X4 6.5mil | 63.5 | 4.95 | 165.1 |
| 9 | WF X4 6.0mil | 76.2 | 10.0 | 152.4 |

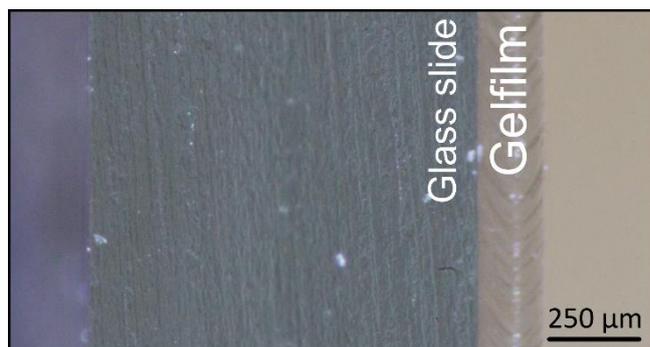

**Figure S3**. Cross-section optical microscopy image to verify the thickness of the Gelfilm strip, which is in good agreement within the experimental resolution with the value provided by the manufacturer (6.5 mil, 165.1 µm).

**Calibration of the optical microscope system**

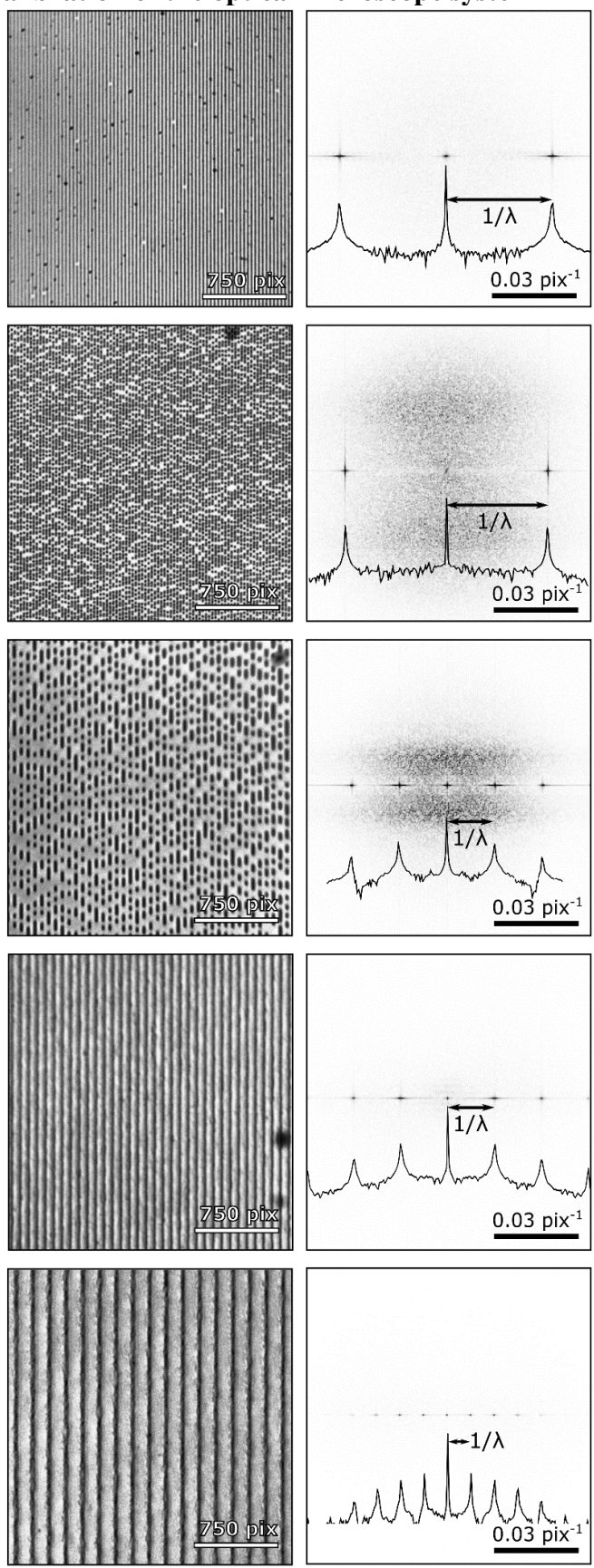

**Figure S4**. Optical microscopy images of a DVD-RW, a DVD, a CD, a 600 lines/mm grating and a 300 lines/mm grating used as reference samples to calibrate the optical microscope. The FFT of the images are shown besides the optical image where the period of the tracks can be easily extracted.

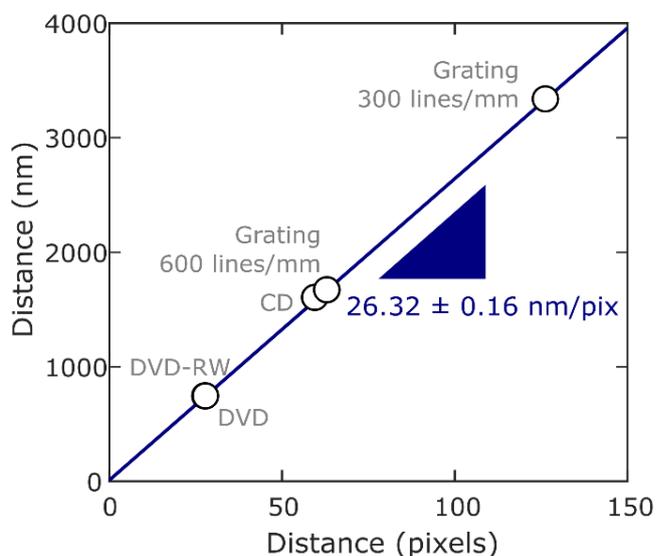

**Figure S5**. Relationship between the measured track distance (in pixels) and the expected values (in nm) for reference samples: DVD-RW, DVD, CD, and optical gratings with 300 lines/mm (Thorlabs GR13-0305) and 600 lines/mm (Thorlabs GR13-0605). The linear fit of the data provides the calibration of the optical microscope system.

**Table S2**. Summary of the reference samples measured to calibrate the optical microscope system.

| Reference sample | Measured track distance (pixels) | Expected track distance (nm) |
|---|---|---|
| DVD | 27.82 | 740 |
| DVD-RW | 28.00 | 740 |
| CD | 59.63 | 1600 |
| 600 lines/mm | 63.14 | 1666.67 |
| 300 lines/mm | 126.42 | 3333.33 |

**Thickness determination**

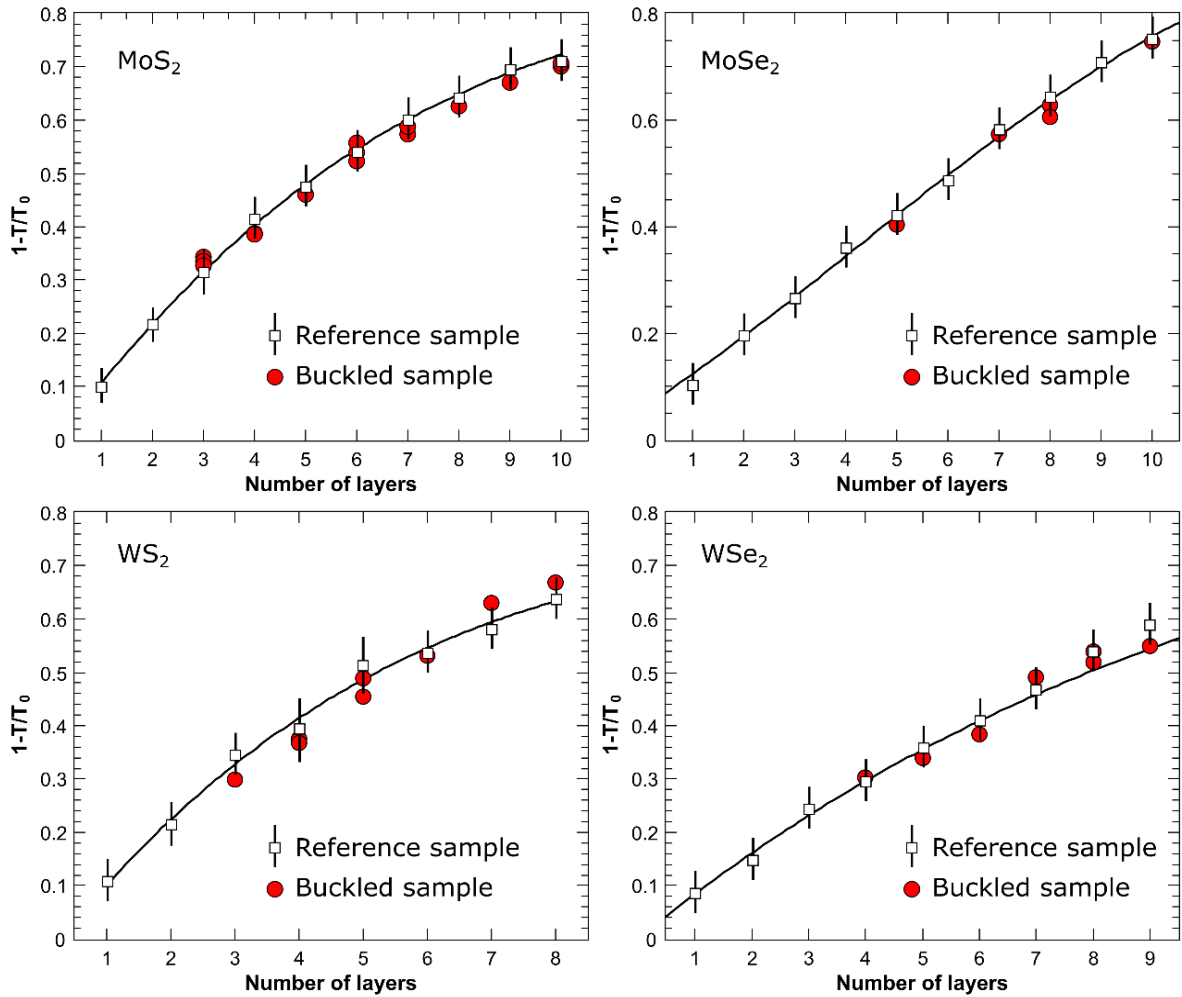

**Figure S6**. Thickness-dependent "absorption" ($1-T/T_0$) of the blue channel, extracted from the transmittance images, as a function of the number of layers. The white squares are the absorption measured on 248 different reference $MoS_2$ flakes, 212 different reference $MoSe_2$ flakes, 212 different reference $WSe_2$ flakes and 223 different reference $WS_2$ flakes with different thicknesses on Gelfilm (not subjected to buckling). The red circles are the data measured on the buckled $MoS_2$, $MoSe_2$, $WS_2$ and $WSe_2$ flakes used to determine the Young's modulus.

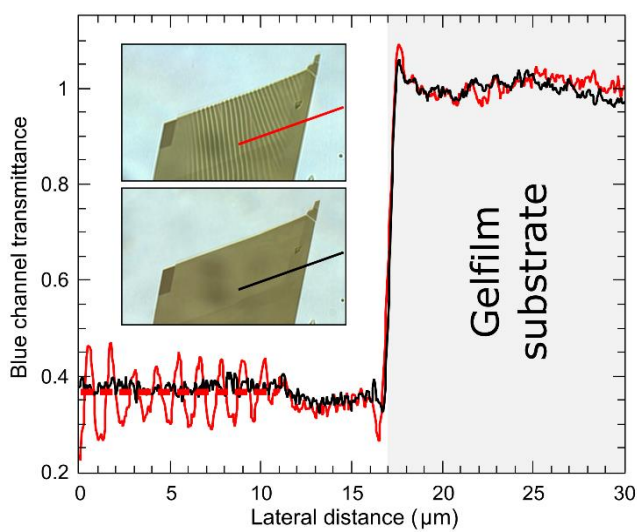

**Figure S7**. Blue channel transmittance ($T/T_0$) extracted from the linecut displayed in the inset for the multilayer $MoS_2$ flake displayed in Figure 1b before and after releasing the strain on the elastomer substrate. The average transmittance value on the rippled region of the sample (after releasing the strain) matches almost perfectly with the value measured on the flat $MoS_2$ before releasing the strain.

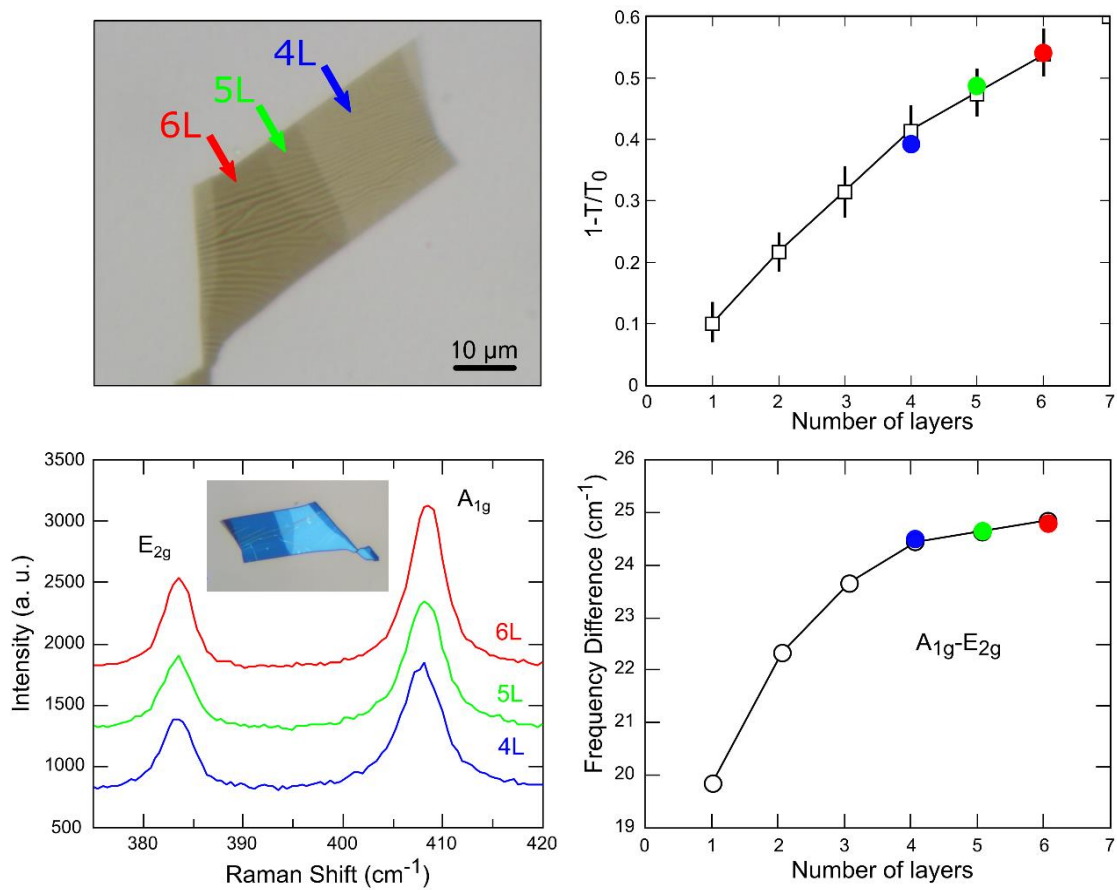

**Figure S8**. Comparison between the thickness determination using the blue channel transmittance and Raman spectroscopy for the same $MoS_2$ flake. (a) Transmission mode optical microscopy image of a multilayer $MoS_2$ flake with regions of different number of layers. (b) Thickness dependent "absorption" ($1-T/T_0$) of the blue channel, extracted from the transmittance image at the highlighted locations, compared with the reference dataset shown in Figure S6. The thickness of these three regions have been determined to be 4L, 5L and 6L. (c) Raman spectra acquired on the same regions after transferring the flake onto a $SiO_2/Si$ surface to iron the ripples, thus avoiding any artefact arising from strain. (d) Raman shift difference between the $A_{1g}$ and $E_{2g}$ modes measured on the same locations where the transmittance was measured. The data has been compared with the thickness-dependent Raman shift difference reported in Ref. [40] in order to determine the thickness of the regions. Their Raman shift difference is compatible with the thickness of 4L, 5L and 6L. Note that the determination of the thickness for flakes thicker than 4L with Raman results way more challenging than with the blue channel transmittance analysis as the Raman shift magnitude starts to saturate at 4-5 layers thick.

**Young's modulus determination for $MoSe_2$, $WS_2$ and $WSe_2$**

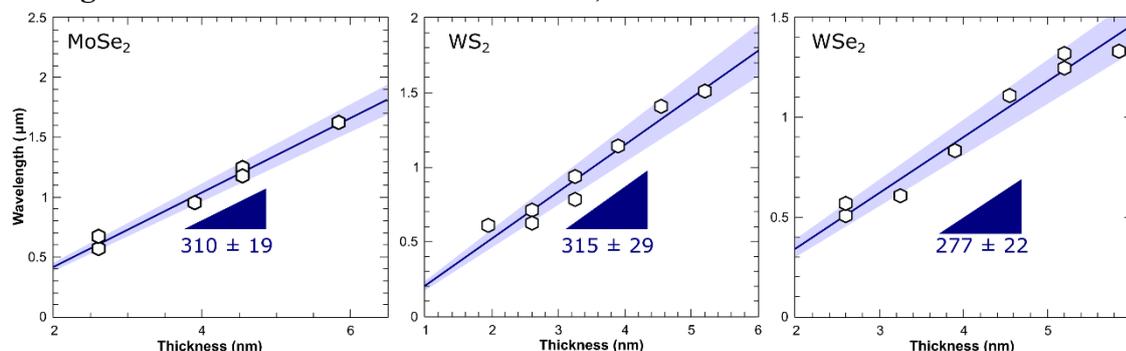

**Figure S9**. Relationship between the wavelength of the ripples and the thickness of the $MoSe_2$, $WS_2$ and $WSe_2$ flakes. The solid dark blue line is a linear fit to the experimental data, the shaded light blue area indicates the uncertainty of the fit. The slope of the wavelength vs. thickness relationship, from which the Young's modulus can be determined, is also included in the plot.

**Table S3**. Summary of the resulting wavelength vs. thickness relationships and the obtained Young's modulus. The value of the Poisson's ratio (and the reference where these values is obtained) used for the Young's modulus calculation has been included as well.

| Material | λ/h | ν | E (GPa) |
| --- | --- | --- | --- |
| $MoS_2$ | 322 ± 15 | 0.27[25] | 246 ± 35 |
| $MoSe_2$ | 310 ± 19 | 0.23[26] | 224 ± 41 |
| $WS_2$ | 315 ± 29 | 0.22[26] | 236 ± 65 |
| $WSe_2$ | 277 ± 22 | 0.19[26] | 163 ± 39 |

**Example of ripples below the microscope resolution**

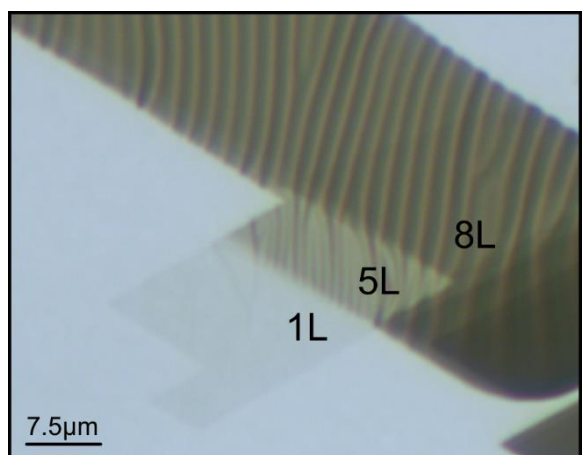

**Figure S10**. Optical microscopy image of a $MoS_2$ flakes with regions of different thicknesses where ripples are clearly visible for the thicker regions. The single-layer region should present ripples with periodicity of ~ 200-300 nm that cannot be resolved with optical microscopy. Therefore, this technique is limited to study flakes thicker than 2-3 layers if an optical microscope is used to determine the ripple period. This limitation could be circumvented using an AFM.

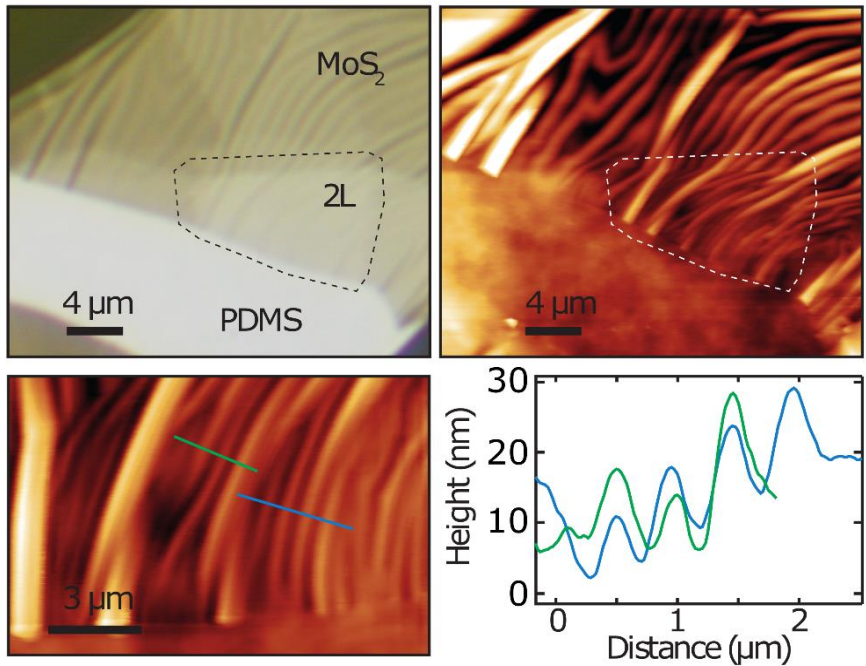

**Figure S11**. (top left) Optical microscopy image of a MoS$_2$ flakes with regions of different thicknesses where ripples are clearly visible for the thicker regions. The bilayer region should present ripples with periodicity of ~400-500 nm that cannot be resolved with optical microscopy. (top right) AFM topography image of the same region illustrating how the bilayer region, in fact, presents ripples that cannot be resolved with the optical microscope. (bottom left) higher magnification topography image in the bilayer region. (bottom right) topographic line profiles where the ripple periodic oscillations (with an amplitude < 5 nm) can be resolved.

**Sensitivity of the determination of the Young's modulus on the choice of Poisson's ratio values**

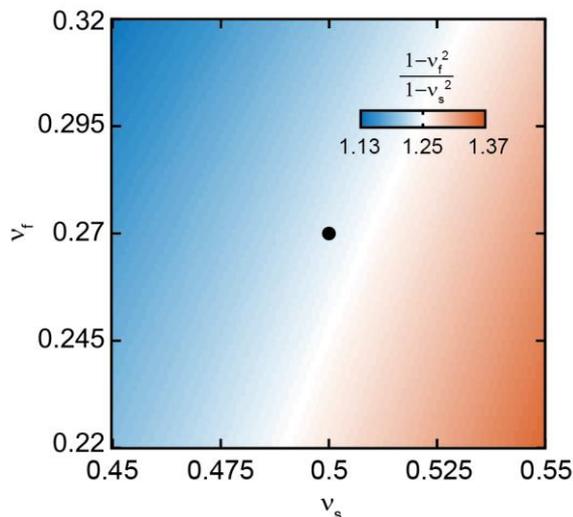

**Figure S12**. Colour plot of the function $Z = \frac{(1-v_f^2)}{(1-v_s^2)}$, which appears in Equation 2 of the main text. The horizontal axis represents the Poisson's ratio of the substrate, the vertical axis the Poisson's ratio of the flake and the different colours the values of the function $Z$. The black dot indicates the choice of and used in the main text for the determination of the Young's modulus of MoS$_2$. The function $Z$ is slowly varying in the region considered.